 \documentclass[cits]{PoS}

\title{Higgs coupling measurements and impact on the MSSM}

\ShortTitle{Higgs coupling measurements and impact on the MSSM}

\author{B\'eranger Dumont%
         \\
        LPSC, Universit\'e Grenoble-Alpes, CNRS/IN2P3, \\ 53 Avenue des Martyrs, F-38026 Grenoble, France\\
        E-mail: \email{dumont@lpsc.in2p3.fr}}

\abstract{Run~I of the LHC has not revealed any sign of new physics beyond the Standard Model (BSM). However, the discovery of an SM-like Higgs boson with mass around 125~GeV opens up new possibilities for probing various BSM scenarios with enlarged Higgs sectors and/or new particles affecting the loop-induced processes or opening new decay modes.
We will present how we derive constraints on new physics from the Higgs measurements performed by the ATLAS and CMS collaborations. 
The impact of these measurements will then be assessed in the context of the general phenomenological Minimal Supersymmetric Standard Model (MSSM) and in the MSSM with a light neutralino as a dark matter candidate.
}

\FullConference{XXII. International Workshop on Deep-Inelastic Scattering and Related Subjects \\
                 28 April - 2 May 2014 \\
                 Warsaw, Poland}

\begin{document}

%%%%%%
\section{Introduction}
%%%%%%

The measurements performed at the LHC on the Higgs boson with a mass of about 125~GeV~\cite{Aad:2012tfa,Chatrchyan:2012ufa,ATLAS-CONF-2014-009,CMS-PAS-HIG-13-005} revealed crucial information on the properties of this new particle beyond its mass. While no sign of a deviation from an SM-like Higgs boson was observed, the various measurements can be used to discriminate between models of new physics in which the Higgs couplings to SM particles are modified and/or additional decay modes open up. After a short discussion on the experimental data, the status of the phenomenological MSSM and of the MSSM with a light neutralino as a dark matter (DM) candidate will be presented in light of the Run~I LHC Higgs results.

%%%%%%
\section{Higgs likelihood}
%%%%%%

\begin{figure}[ht]\centering
\includegraphics[scale=0.39]{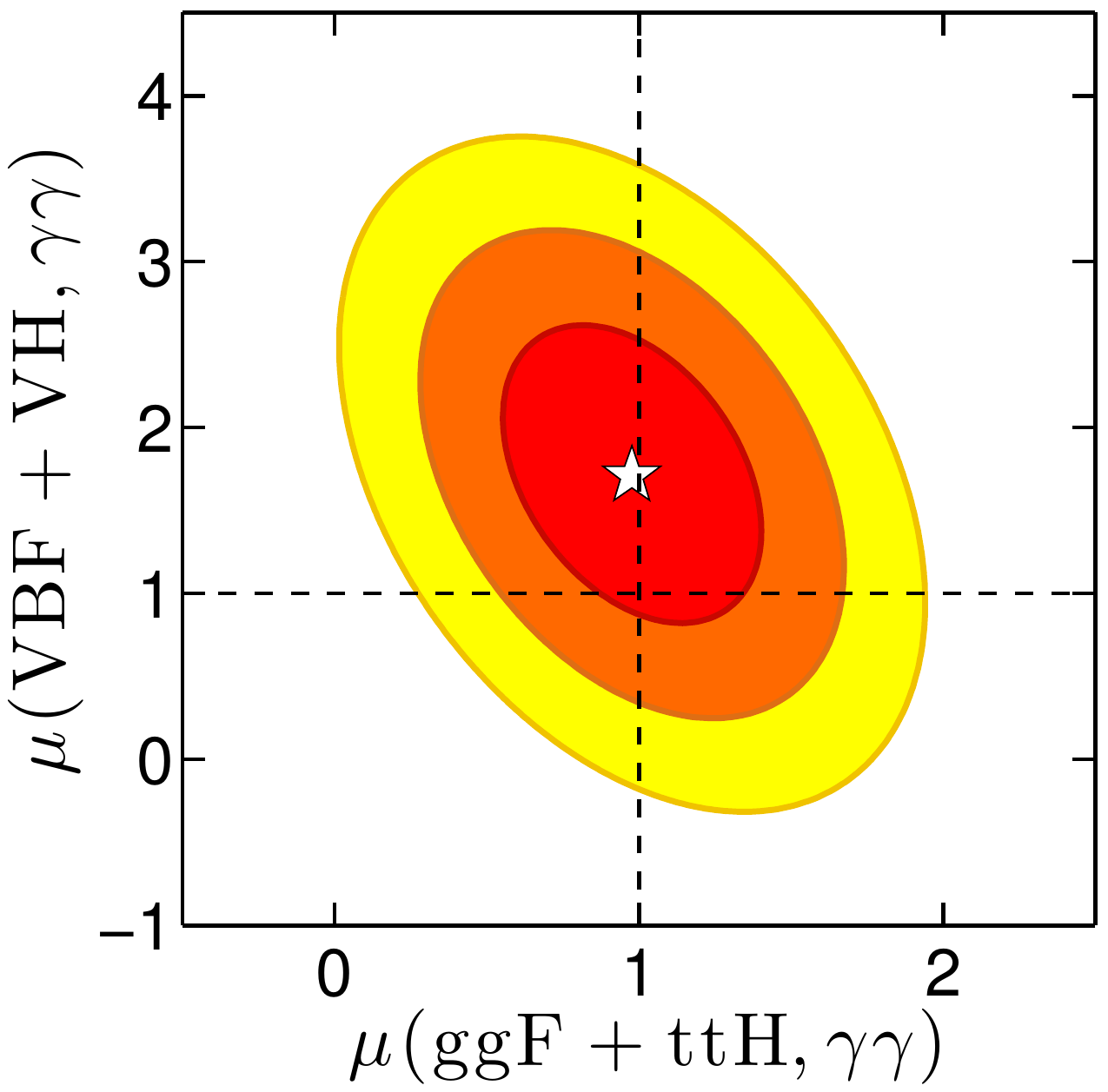} \quad \ 
\includegraphics[scale=0.39]{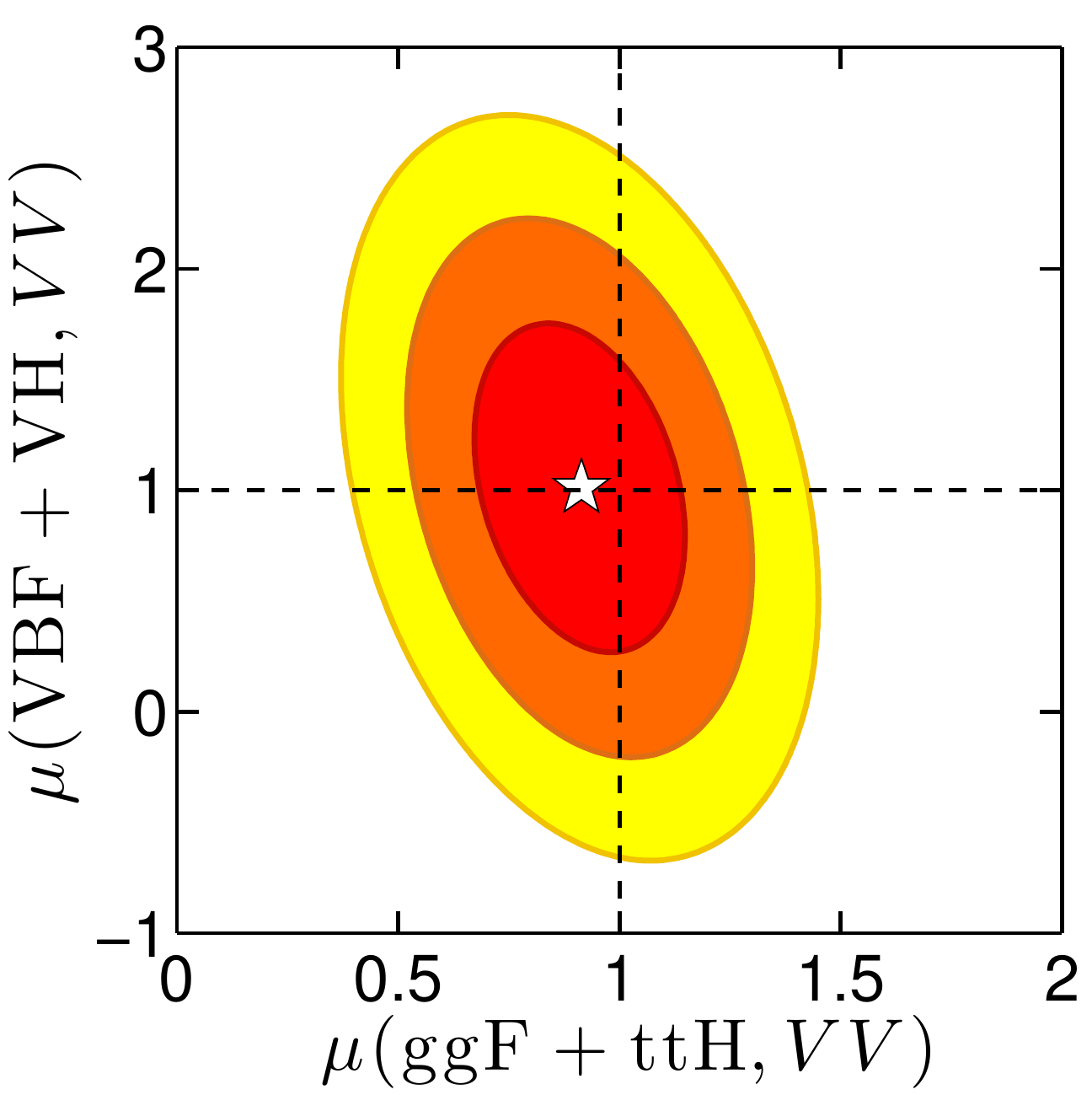}

\includegraphics[scale=0.4]{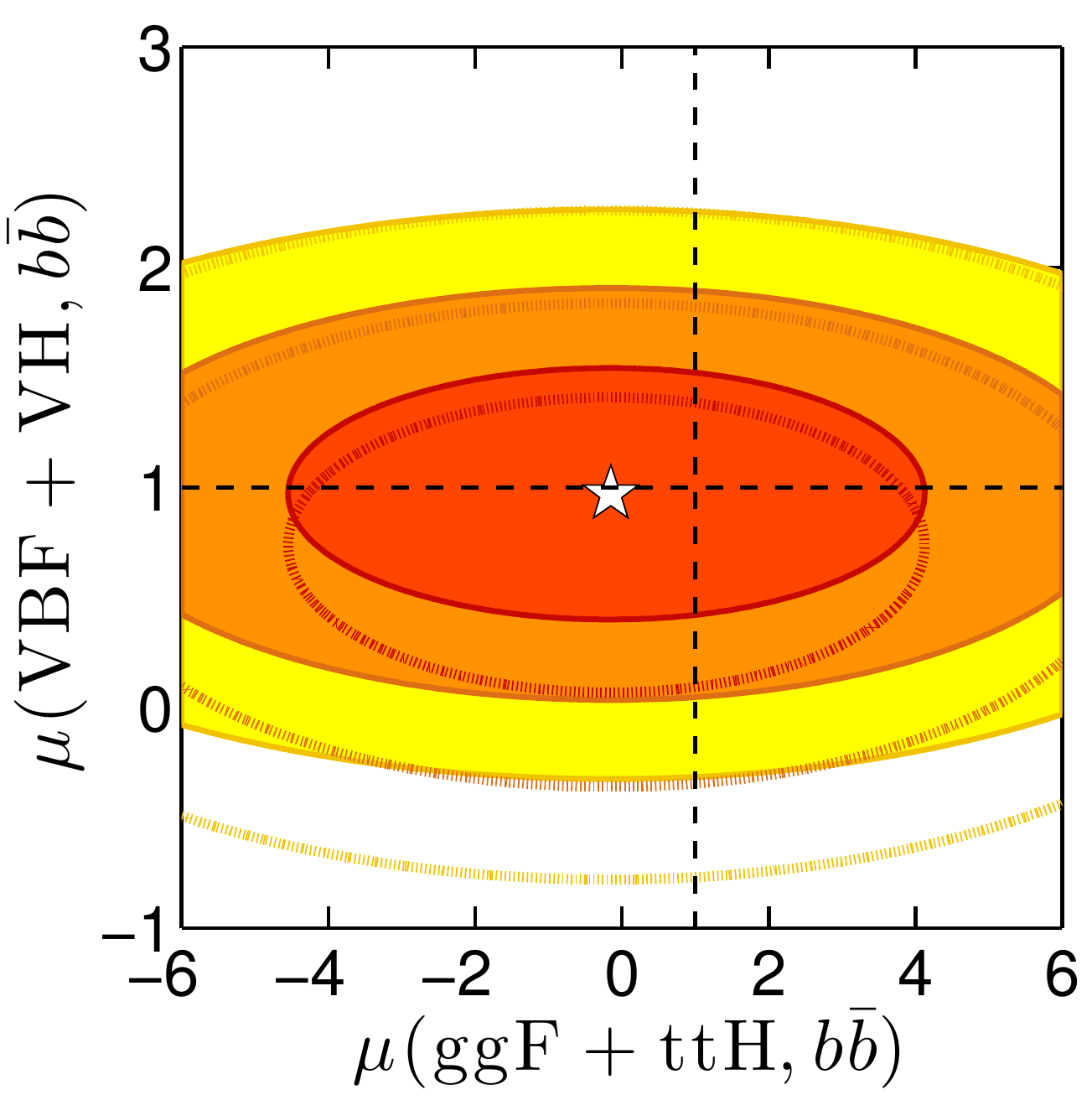}\quad
\includegraphics[scale=0.4]{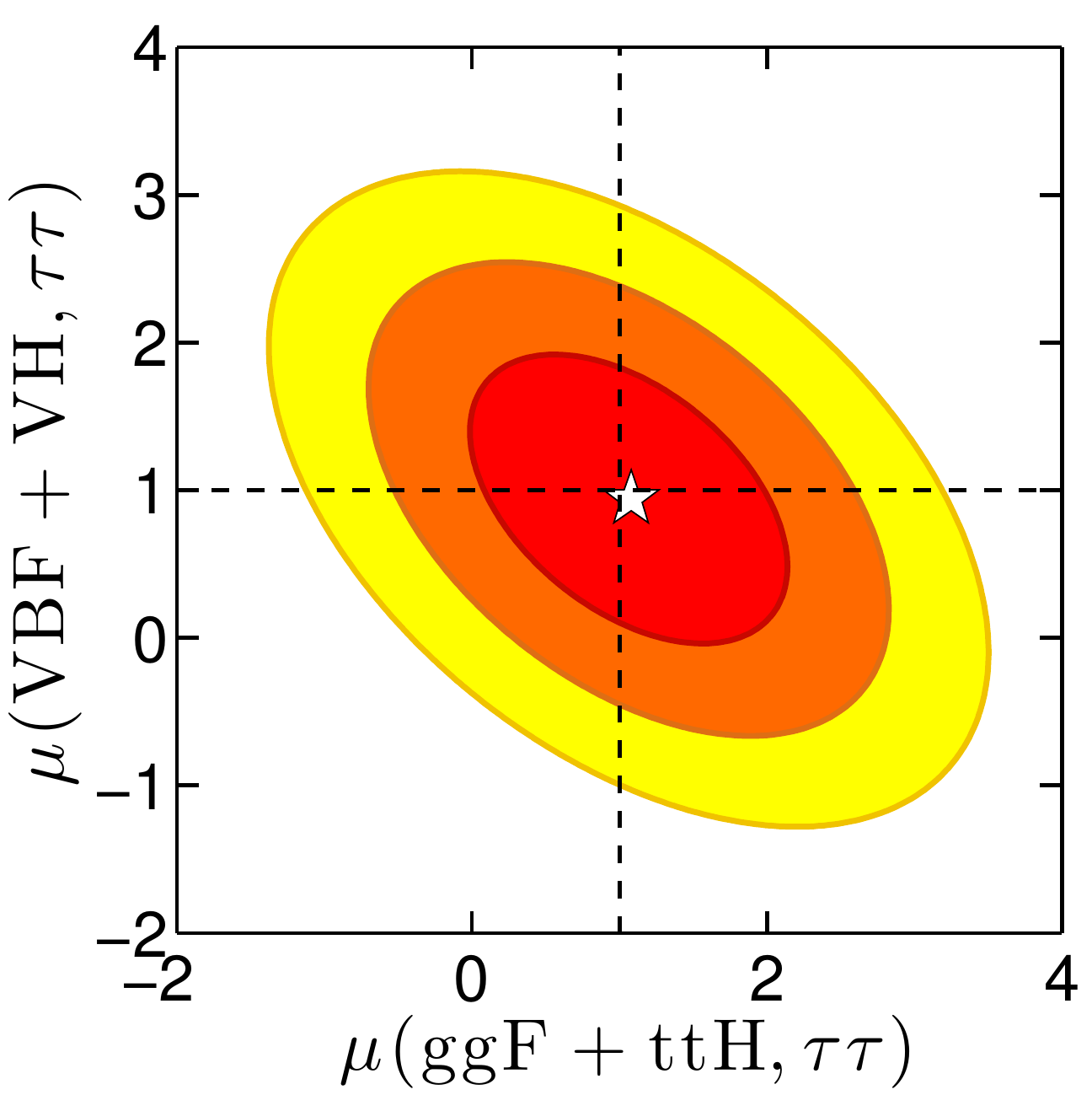}
\caption{Signal strength ellipses for the $\gamma\gamma$, $VV=ZZ,WW$, $b\bar b$ and $\tau\tau$ channels, using all results up to the 2013 LHCP conference as was done in~\cite{Belanger:2013xza}. The filled red, orange and yellow ellipses show the 68\%, 95\% and 99.7\%  CL regions, respectively, derived by combining the ATLAS, CMS and Tevatron results. The red, orange and yellow line contours in the bottom left plot show how these ellipses change  when neglecting the Tevatron results. 
\label{fig:ellipses} }
\end{figure}

In order to determine as precisely as possible the properties of the 125~GeV Higgs boson, a large number of measurements were carried out at the LHC with the data collected during Run~I. The results, given in terms of signal strengths $\mu = \sigma / \sigma_{\rm SM}$, have correlated systematics uncertainties which make it difficult to constrain models of new physics in a precise way from outside the experimental collaboration. Fortunately, the ATLAS and CMS collaborations now systematically present results in the $(\mu({\rm ggF+ttH}, Y), \mu({\rm VBF+VH}, Y))$ plane~\cite{ATLAS-CONF-2014-009,CMS-PAS-HIG-13-005}, where the five production modes of the SM are grouped into just two effective modes (gluon fusion $+$ associated production with top quarks, ${\rm ggF+ttH}$, and vector boson fusion $+$ associated production with a vector boson, ${\rm VBF+VH}$) and where $Y$ are the decay modes of the SM Higgs boson (currently $\gamma\gamma$, $ZZ^*$, $WW^*$, $b\bar b$, and $\tau\tau$). This information can be used directly to constrain a very wide class of new physics models, as was discussed in detail in~\cite{Belanger:2013xza}. The constraints from ATLAS, CMS and the Tevatron can then be combined and the resulting likelihood expressed with a simple $\chi^2$ formula, as was done in~\cite{Belanger:2013xza} using all experimental results up to the LHCP 2013 conference. These constraints can be depicted as combined signal strengths ellipses for each of the final states $Y$, as is shown in Fig.~\ref{fig:ellipses}, and will be used to constrain the MSSM in the next sections.
A public tool for fitting the Higgs likelihood, based on this approach, and deriving constraints on a broad class of new physics models using all available experimental data beyond the Gaussian approximation is in preparation~\cite{Lilith}.

%%%%%%
\section{Impact on the phenomenological MSSM}
%%%%%%

The phenomenological MSSM (pMSSM) is a 19-dimensional parametrization of the weak-scale Lagrangian of the MSSM~\cite{Djouadi:1998di}.
It captures most of the phenomenological 
features of the R-parity conserving MSSM and, most importantly, encompasses and goes beyond a broad 
range of more constrained SUSY models.
We examined the status of the pMSSM in light of the experimental results on the Higgs boson in~\cite{Dumont:2013npa}. In this work, a Bayesian approach was taken and we investigated how the latest LHC results on the properties of the 125~GeV Higgs state impact the probability distributions of the pMSSM parameters, masses and other observables, using the likelihood presented in Section~1 on top of other relevant experimental constraints.

We found that significant deviations of the Higgs signal strengths from 1 are still allowed, even when $m_A$ and the masses of all SUSY partners are above 1~TeV. This is mostly coming from the SUSY radiative corrections to the bottom Yukawa coupling~\cite{Carena:1999py,Eberl:1999he}, which can be large for large $\tan \beta$. To a good approximation, the corrections reads
\begin{eqnarray}
   \Delta_b \equiv \frac{\Delta m_b}{m_b} \simeq 
   \left[ \frac{2\alpha_s}{3\pi} \mu m_{\tilde{g}}\, I(m_{\tilde{g}}^2, m_{\tilde{b}_1}^2, m_{\tilde{b}_2}^2) +
            \frac{\lambda_t^2}{16\pi^2} A_t \mu \, I(\mu^2,m_{\tilde{t}_1}^2, m_{\tilde{t}_2}^2) \right] \tan\beta \,, \label{pmssm-Deltamb} 
\end{eqnarray}
where $I(x,y,z)$ is of order $1/{\rm max}(x,y,z)$~\cite{Djouadi:2005gj}. Since $H_{\rm SM} \to b\bar b$ is the dominant decay mode of the SM Higgs at 125~GeV, a non-vanishing $\Delta_b$ will impact all signal strengths via the total width.

\begin{figure}[ht]\centering
\includegraphics[width=0.40\textwidth]{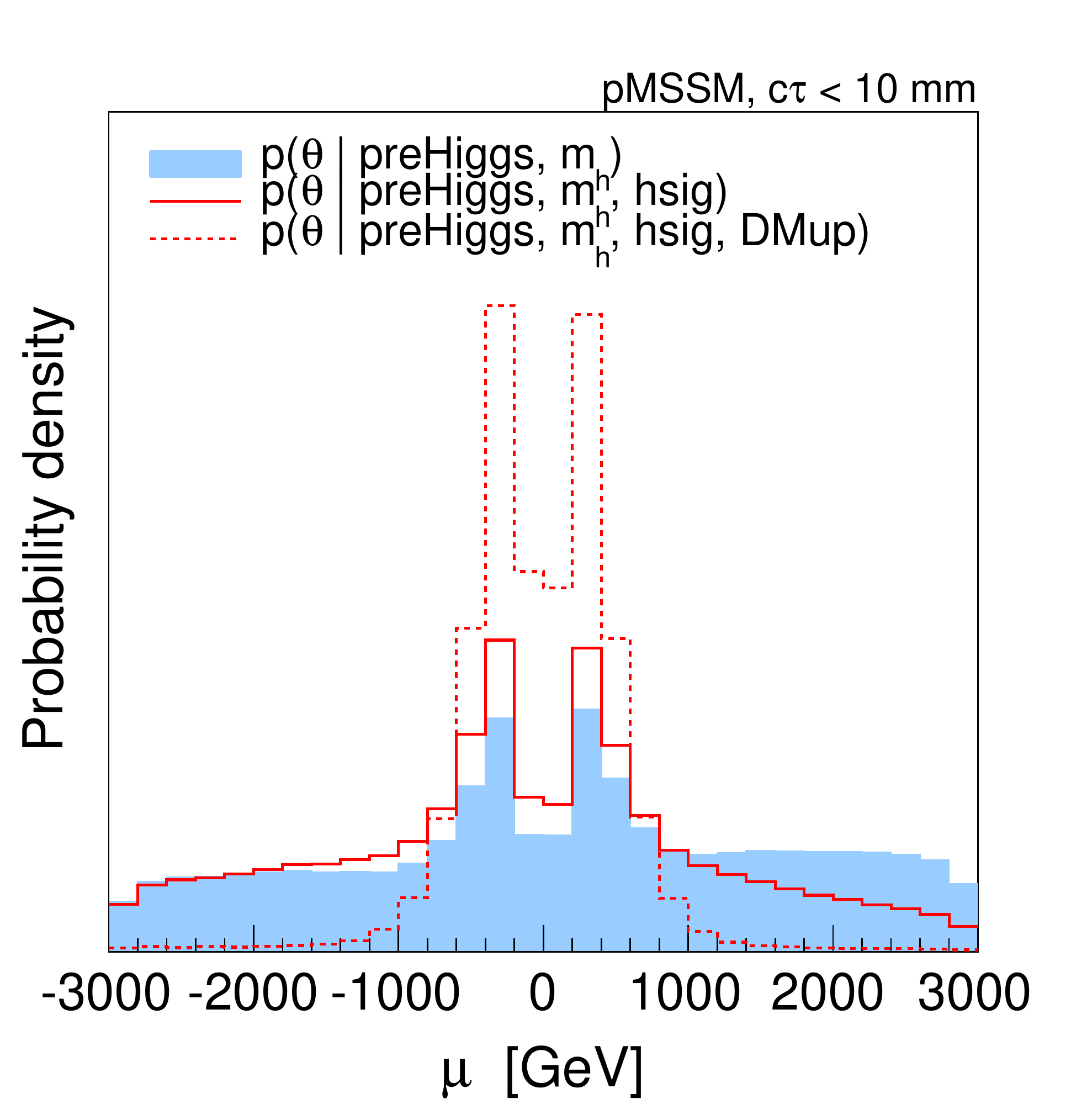}
\includegraphics[width=0.40\textwidth]{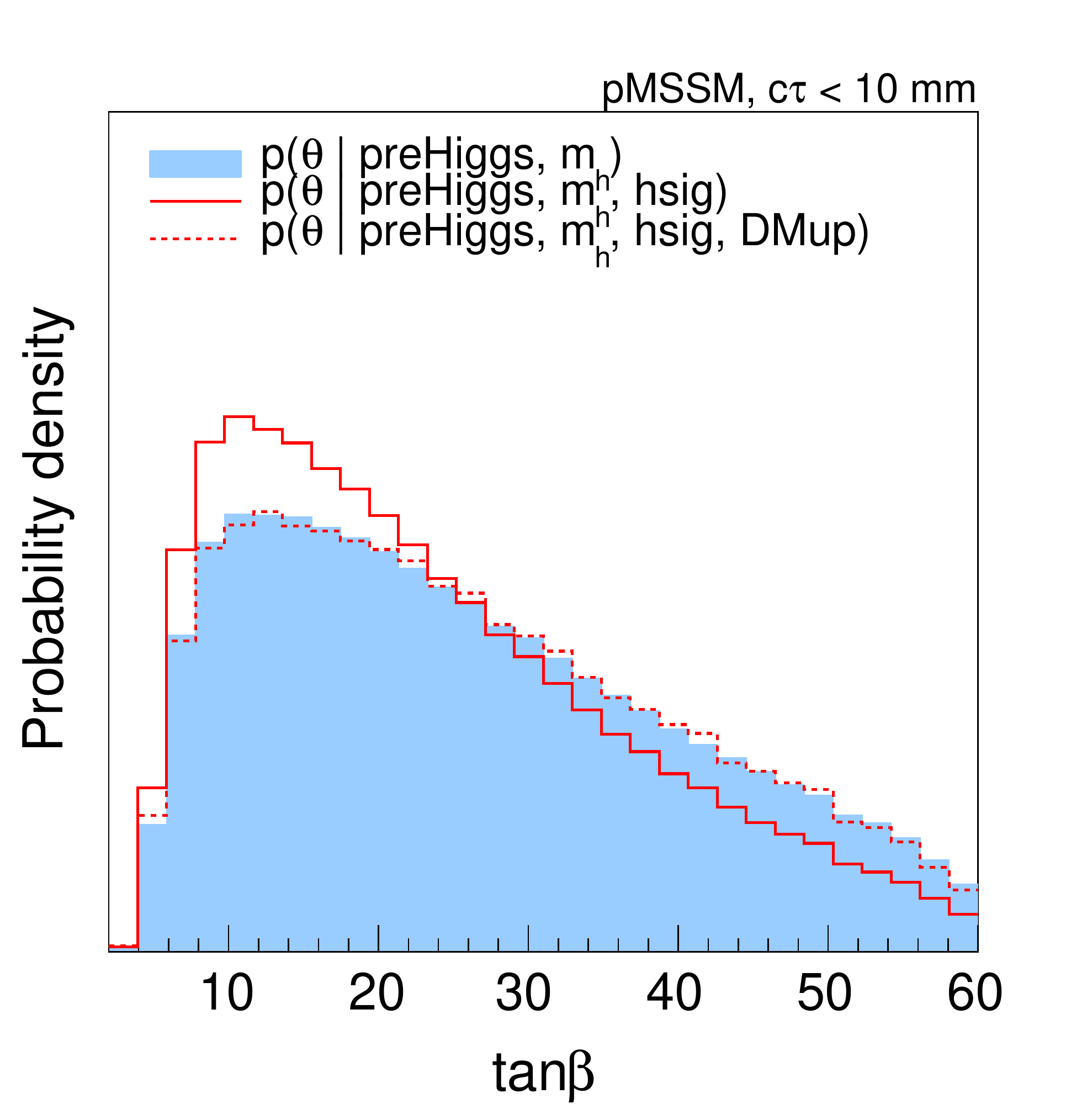}
\caption{Marginalized 1D posterior densities for the parameters $\mu$ (left) and $\tan \beta$ (right), showing the effect of the 
Higgs signal strength measurements. 
The light blue histograms show the distributions  based on the ``preHiggs'' measurements 
plus requiring in addition $m_h\in [123,\,128]$~GeV.
The solid red lines, labelled ``hsig'', are the distributions when moreover taking into account 
the measured Higgs signal strengths in the various channels. The limits from searches for the heavy Higgses ($H$ and $A$) 
are also included in the red line-histograms, but have a negligible effect. 
The dashed red lines, labelled ``DMup'', include in addition an upper limit on the 
neutralino relic density and the recent direct DM detection limit from LUX~\cite{Akerib:2013tjd}.
\label{fig:pmssm} }
\end{figure}

We found that the current LHC results on the Higgs boson have a significant impact on the posterior distributions of $\mu$ and $\tan\beta$, as illustrated in Fig.~\ref{fig:pmssm}. Here, ``preHiggs'' measurements include flavor and low-energy constraints as well as the limits on the masses of SUSY particles from LEP. Moderate values of $\tan \beta$ are preferred to suppress the $\Delta_b$ correction, and large positive values of $\mu$ are disfavored because it suppresses all signal strengths while a slight excess is observed in the $H \to \gamma\gamma$ channels in ATLAS.
We have not taken into account the recent LHC limits from direct SUSY searches, but we checked that 
our conclusions do not change when requiring gluino and squark masses above 1~TeV. The conclusions drawn 
from the Higgs sector are thus orthogonal to those from the SUSY searches. 
In particular, 
this makes our results directly comparable to the pMSSM interpretation of the CMS SUSY searches at 7--8~TeV~\cite{CMSpMSSM}.

%%%%%%
\section{Impact on the MSSM with light neutralino dark matter}
%%%%%%

It is also interesting to focus on a more specific scenario within the general pMSSM, where the neutralino is as light as possible and a viable DM candidate. This was done in~\cite{Belanger:2013pna}, taking into account the results of the SUSY searches performed at the LHC using {\tt SModelS}~\cite{Kraml:2013mwa} in addition to the constraints already discussed in Section~3. For the Higgs constraints, we made use of the signal strength ellipses presented in Fig.~\ref{fig:ellipses}.  A given point in parameter space was considered as excluded if one of these four 2D signals strengths fell outside the 95\%~CL experimental region.

\begin{figure}[ht]\centering
\includegraphics[width=0.40\textwidth]{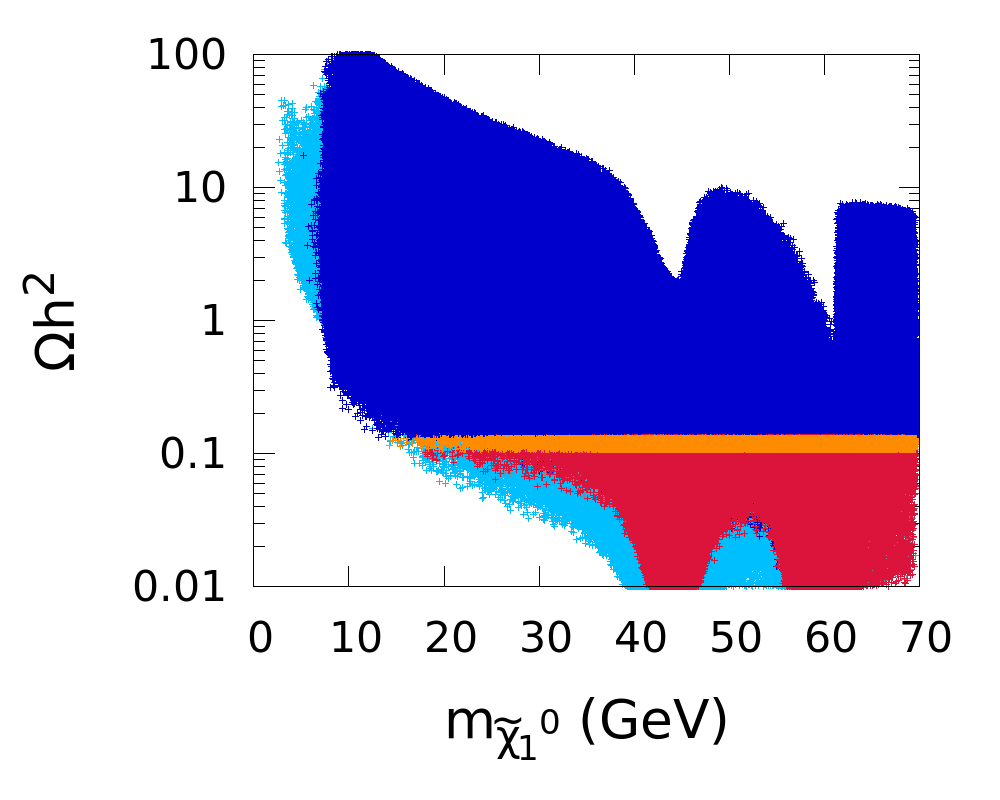}
\includegraphics[width=0.40\textwidth]{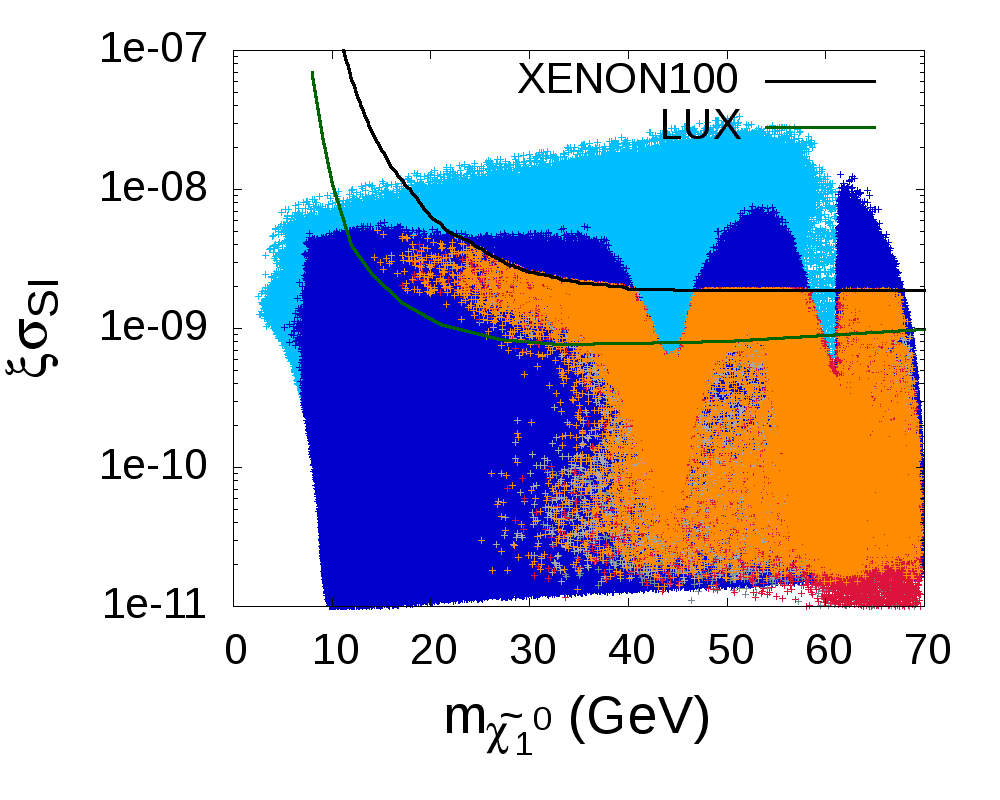}
\caption{Relic density $\Omega h^2$ (left) and rescaled spin independent scattering cross section $\xi \sigma_{\rm SI}$ (right) as function of the neutralino mass, with $\xi = \Omega h^2/0.1189$.
Cyan points  fulfill the ``basic constraints'' and also pass 
the limit on $A^0, H^0 \to \tau^+ \tau^-$ from CMS; 
blue points are in addition compatible at 95\%~CL with all Higgs signal strengths based on the global fit.
Finally,
red (orange) points obey also the relic density constraint $\Omega h^2<0.131$ ($0.107<\Omega h^2<0.131$) 
and abide the direct detection limits from XENON100 on $\sigma_{\rm SI}$. The 2013 limit from LUX~\cite{Akerib:2013tjd}, which came out after the publication of~\cite{Belanger:2013pna}, is shown as a green line. 
\label{fig:lightdm} }
\end{figure}

\begin{figure}[ht]\centering
\includegraphics[width=0.42\textwidth]{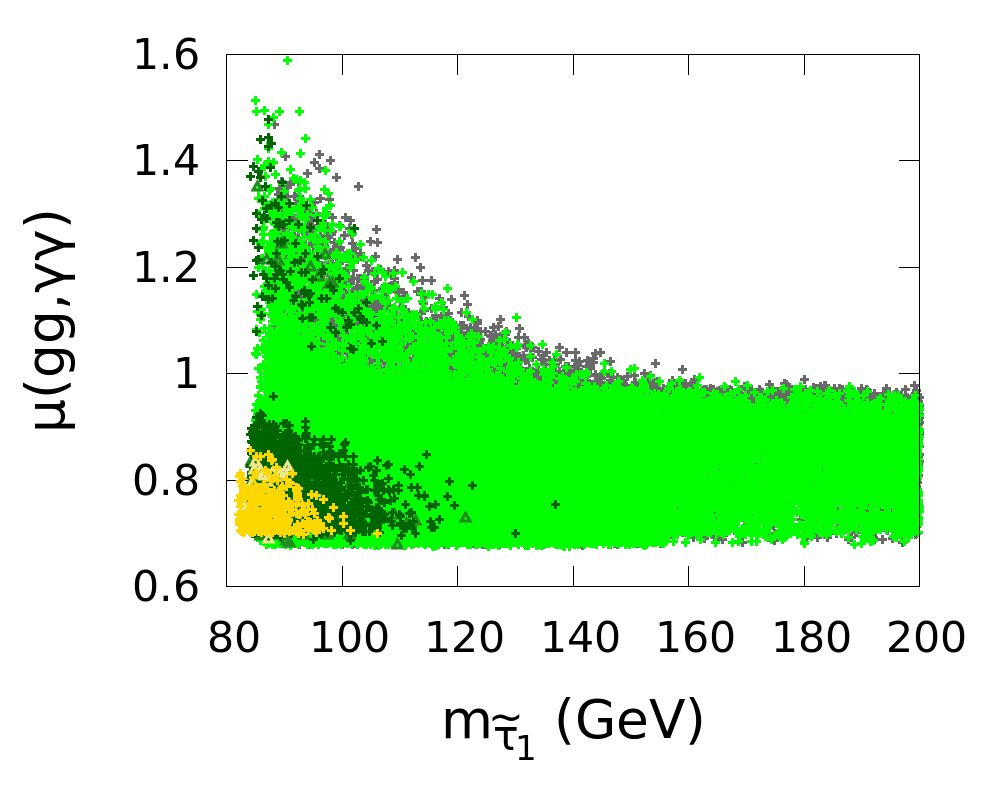}
\caption{Implications of the light neutralino DM scenario for Higgs signal strengths for points passing all constraints. The yellow, dark green, light green and gray points have $\tilde\chi^0_1$ masses of 15--25~GeV, 25--35~GeV, 
35--50~GeV and 50--60~GeV, respectively. For more details, see~\cite{Belanger:2013pna}. 
\label{fig:lightgam} }
\end{figure}

The results are as follows. First, we found that the upper bound on the relic density of $\Omega h^2 \lesssim 0.11$ sets a lower bound on the neutralino mass of about 15~GeV, as can be seen in the left plot of Fig.~\ref{fig:lightdm}. These very light neutralinos are a mixture of bino and higgsino (implying a light chargino), and come with rather light staus in order to maximize the annihilation of DM in the early universe. This region is however put under strong pressure by the latest limits on the direct direction of DM, see the right plot of Fig.~\ref{fig:lightdm} (in all plots, points are retained if they pass the XENON100 90\%~CL bound).
The impact on the Higgs signal strengths is shown in Fig.~\ref{fig:lightgam}. While very light DM with mass below 25~GeV always leads to $\mu(gg, \gamma\gamma) < 1$ because of the decay $h^0 \to \tilde\chi^0_1 \tilde\chi^0_1$, neutralinos with masses in the 25--35 GeV range may also lead to an excess in $\mu(gg, \gamma\gamma)$ if staus are light and maximally mixed~\cite{Carena:2012gp}. These deviations represent a promising way of probing light neutralino scenarios during Run~II of the LHC.

%%%%%%
\section{Conclusions}
%%%%%%

The measurements of the properties of the Higgs boson at Run~I of the LHC already constrain BSM physics, beyond the information on the mass of the new boson. It is possible to encapsulate the constraints into a simple likelihood which can then be used to constrain generic modifications to the strengths of the Higgs couplings. We discussed the impact on the general phenomenological MSSM and on a subset of the pMSSM with a light neutralino as a dark matter candidate. We found that the current measurements already provide quite relevant constraints and pointed out promising prospects for probing the parameter space of the MSSM with Run~II LHC data from 2015 onwards.

%%%%%%
\acknowledgments{This talk was based on fruitful collaboration with Gevevi\`eve Bélanger, Guillaume Drieu La Rochelle, Ulrich Ellwanger, Rohini~M.\ Godbole, John F.\ Gunion, Sabine Kraml, Suchita Kulkarni and Sezen Sekmen. I would
moreover like to thank the conveners of WG3 for the invitation to
this conference. The work presented here was supported in part by IN2P3 under contract
PICS FR-USA No. 5872 and by the French ANR DMAstroLHC.}
%%%%%%

\end{document}